\documentclass{ptephy_v1}%%%%%% to generate preprint number with ptep logo

\preprintnumber{XXXX-XXXX} %%% %%% Insert preprint number here

\newcommand{\bea}{\begin{eqnarray}}
\newcommand{\eea}{\end{eqnarray}}
\newcommand{\be}{\begin{equation}}
\newcommand{\ee}{\end{equation}}

\newcommand{\gre}[1]{\textcolor[rgb]{0.0,0.5,0.0}{\bf #1}}

\def\be{\begin{equation}}
\def\ee{\end{equation}}
\def\bea{\begin{eqnarray}}
\def\eea{\end{eqnarray}}

\usepackage{cancel}

\begin{document}

\title{Residual Tensor Force Effects on the Gamow-Teller states in Magic Nuclei, $^{48}$Ca, $^{90}$Zr, $^{132}$Sn, and $^{208}$Pb }

%%%% To generate auto affiliation numbers please use \author{}\affil{} command

\author{Eunja Ha \footnote{ejaha@hanyang.ac.kr}}
\affil{Department of Physics and Research Institute for Natural Science, Hanyang University, Seoul 04763, Korea}
\author{Myung-Ki Cheoun \footnote{cheoun@ssu.ac.kr (Corresponding Author)}}
\affil{Department of Physics and Origin of Matter and Evolution of Galaxies (OMEG) Institute, Soongsil University, Seoul 156-743, Korea}
\author{H. Sagawa \footnote{sagawa@ribf.riken.jp}}
\affil{RIKEN, Nishina Center for Accelerator-Based Science, Wako 351-0198, Japan and Center for Mathematics and Physics, University of Aizu, Aizu-Wakamatsu, Fukushima 965-8560, Japan}

%%% To include the collaborator name... Please use the command "\collaborator"
%%% For example: \collaborator{ATLAS Collaboration}

\begin{abstract}%
We investigate the tensor force (TF) effect %in the residual interaction
 on the Gamow-Teller (GT) transitions in four magic nuclei, $^{48}$Ca, $^{90}$Zr, $^{132}$Sn and $^{208}$Pb. The TF is taken into account by using the Br\"uckner $G$-matrix theory with the charge-dependent (CD) Bonn potential  as the residual interaction of charge-exchange quasiparticle random phase approximation (QRPA).
 We found that particle-particle ($p-p$) tensor interaction does not affect the GT transitions because of the closed shell nature in the nuclei, but repulsive particle-hole ($p-h$) residual interaction for the  $p-h$ configuration of spin-orbit partners dominates the high-lying giant GT states for all of the nuclei.
It is also shown that appreciable GT strengths are shifted to lower energy region by the attractive $p-h$ TF for the same  $j_\pi=j_\nu$ configuration, and produce the low-lying GT peak about 2.5 MeV  in $^{48}$Ca. Simultaneously, in $^{90}$Zr and $^{132}$Sn, the low-energy strength appears as a lower energy shoulder near the main GT peak. On the other hand, the shift of the low-lying GT state is not seen clearly for $^{208}$Pb because of the strong spin-orbit splitting of high $j$ orbits, which dominates the GT strength.
\end{abstract}

\subjectindex{Tensor Force, Gamow-Teller Transition, Pairing Matrix Elements, Br\"uckner $G$-matrix}

\maketitle

\section{Introduction}
The tensor force (TF) in the nucleon-nucleon ($N$-$N$) interaction is one of key ingredients for understanding various nuclear correlations in finite nuclei.  As already indicated by H. Bethe 1940 \cite{Bethe1940}, and by E. Gerjuoy and J. Schwinger \cite{Schwinger1942}, the deuteron, $(J^{\pi} = 1^+, T=0)$ never been a bound state without the TF, contrary to the 
unbound  di-neutron system $(J^{\pi} = 0^+, T=0)$ in which TF is not active \cite{BW52}. It means that the primordial nucleosynthesis could not be started without the TF. Lots of phase shift analyses of
the $N$-$N$ scattering data show that the TF in the triplet-even channels, $^{3}S_1$ and $^{3}D_1$ states, are strongly attractive \cite{Ericson,Byst1987}. Since the monumental papers regarding the TF, the TF effects in nuclei have been examined in the mean field model from 1970's \cite{Stancu77}, and have been extensively studied by many theoretical models and experiments during last twenty years \cite{Otsuka05,Lesinski2007,Colo2007,Brink2007,Zou2007,Bender2009,
Otsuka2010,Urata17,Bernard16,Suzuki16,Nakada2020,Sagawa14-2,Tani13}.
Specifically, a recent study of the role of the TF by the advanced shell model clarified that the non-central tensor interaction is attractive between $j_<= l -{1/2}$ and $j_> = l + 1/2$ states, but it is repulsive for $j_>$ and $j_>$ states or  for $j_<$ and $j_<$ states \cite{Otsuka05,Otsuka2010,Otsuka2020,Tani13}. The TF effect in the Skyrme force type  has also been discussed in detail in Refs. \cite{Mich2003,Sagawa16,Brink2007,Sagawa14,Lesin2007,Bender2009,Hell2012}, which confirms the TF properties  discussed by the shell models and emphasizes the important role of TF in the density functional theory (DFT).

%\red{The TF effect in nuclei has been extensively studied by many theoretical models and experiments during the last twenty years \cite{Otsuka2020,}.
%Recent study by the advanced shell models regarding the roles of the TF in nuclei shows that the non-central tensor interaction is attractive between neutron and proton pair in $j_<= l -{1/2}$ and $j_> = l + 1/2$ states, but it is repulsive for the case between $j_>$ ($j_<$) and $j_>$ ($j_<$) states \cite{Otsuka05,}. The TF effect is also detailed in the effective Skyrme force type in Refs. \cite{Sagawa14,Sagawa16}, which confirmed the TF properties inside nuclei as discussed by the shell model and emphasized the necessity of the TF study in the density functional theory (DFT).}

In fact, the TF has a non-spherical property and a spin-triplet feature. Therefore, one may conjecture that the nuclear deformation could be closely associated with the TF. It could be a challenging task to find a realistic relation or correlation between the microscopic TF and the macroscopic (or collective) properties like the deformation parameter.
Along this line, recent works in Refs. \cite{Bernard16,Co2021,Suzuki16,Nakada14} discussed the TF effects on the nuclear deformation: they argued
that the TF effects may change the deformation in the Hartree-Fock (HF) Bardeen-Cooper-Schriffer (BCS) scheme with the \cancel{finite} Gogny-type finite-range pairing interaction \cite{Bernard16,Co2021,Otsuka2006} and also by the HF model with M3Y-P6 interaction \cite{Suzuki16,Nakada14}.

The spin-triplet aspect of the TF is also interesting because the isoscalar (IS) pairing has the spin-triplet nature in the neutron-proton channel.  This point  was argued especially  in 
medium-heavy $N \cong Z$ nuclei, where the spin-orbit (SO) coupling could be small and does not overwhelm the spin-triplet IS pairing \cite{Bertsch2010,Bertsch2011}. The microscopic relationship of the TF and the spin-triplet interaction in the mean field still needs more study from the microscopic point of view.

In general, the experimental data of Gamow-Teller (GT) strength have observed the GT Giant Resonance (GTGR) around $E_{GT} - E_F$ = (26 $A^{-1/3} - 18.5 (N -Z)/A )~$MeV with respect to the central peak of the Fermi-type  transition $E_F$ to the isobaric analogue state (IAS) \cite{Naka82,Luto2011}.   The excitation energy GTGR depends on mainly  %mainly for the open shell nuclei 
 induced by the SO force in the mean field,  and is  pushed up further by the repulsive particle-hole ($p-h$) interaction.
 
For the nuclei near closed shells, specifically for $N = Z+2$ nuclei, some GT strengths are shifted to low-energy region mainly by the attractive particle-particle ($p-p$) interaction originating from the TF according to our previous QRPA calculation \cite{Ha2022}.
For example, the experimental data of GT strength distribution for $^{42}Ca(^{3}$He,$t)^{42}$Sc \cite{Fujita15} exhibit the most prominent peak at much lower energy region than the systematically observed GTGR energy. This peak is termed as Low energy Super GT (LeSGT) \cite{Fujita20}, which was attributed to  the attractive property of the $p-p$ interaction  in the low energy GT state of $N=Z+2$ nuclei \cite{Bai-ISP-2014}. Namely, a typical  GTGR excitation between the particle-hole ($p-h$) states of  the spin-orbit (SO) partner is moved to the lower energy region by the attractive $p-p$ interaction.

Here we remind that the IS pairing 
is dominated by the spin-triplet channel, which effectively accounts for most of the neutron-proton ($np$) TF interaction, as a zero range contact interaction. This feature is originated to the fact that the spin-triplet, $^{3}S_1$ and $^{3}D_1$, interaction is larger than that of the spin-singlet $np$ contribution, for example, $^{1}P_1$ interaction.
The TF effects clearly give rise to the similar effect to that by the IS pairing,  as shown in our previous paper \cite{Ha2022}. We also note that the TF is  partially included in the isovector (IV) spin-triplet $np$ pairing part, which is dominated by the attractive spin-triplet interaction ($^{3}P_2$ and $^{3}P_0$), but the magnitude is smaller than that by the IS spin-triplet one.

However, even for some double magic or single magic nuclei, we may find a population of GT states in low-lying excitation as shown for $^{48}$Ca in Fig.\ref{fig1}(a). Such populations in low-lying states are peculiar for $N=Z+2$ nuclei.
In this work, we investigate the energy shift of the GT states %and argue that the shift in the magic nuclei is also
induced by the TF for closed-shell nuclei. This shift  could be an interesting phenomenon because in the closed-shell nuclei the $p-h$ interaction is dominant in contract to $N=Z+2$ nuclei where the $p-p$ interaction plays an important role. In Sec. II, we shortly explain the formulas used in present calculations. Numerical results are presented with detailed discussions in Sec. III. Summary and conclusions are given in Sec. IV. 

\section{Basic Formulas}

Since we study the GT transition in $N>Z$ nuclei, in this wok, we focus on the unlike-pairing correlations, such as neutron-proton pairing ($np)$, as well as the like-pairing (neutron-neutron ($nn)$ and proton-proton $(pp)$ pairing) via the TF by taking Goswami formalism \cite{Gos65,Goodmann70,Goodmann72,Goodmann80} in a BCS approach. Since the theoretical framework for the BCS had been already discussed in details in our previous papers \cite{Ha2022,Ha18-1,Ha18-2,Ha15}, we only present the
basic formula. We start from the following BCS transformation between the quasiparticle and 
the real particle in an $\alpha$ state 
\begin{equation} \label{eq:HFB_1}
\left( \begin{array}{c} a_{1}^{\dagger} \\
  a_{2}^{\dagger} \\
  a_{\bar{1}} \\
  a_{\bar{2}}
  \end{array}\right)_{\alpha} =
\left( \begin{array}{cccc}
u_{1p} & u_{1n} & v_{1p} & v_{1n} \\
u_{2p} & u_{2n} & v_{2p} & v_{2n} \\
-v_{1p} & -v_{1n} & u_{1p} & u_{1n} \\
-v_{2p} & -v_{2n} & u_{2p} & u_{2n}
  \end{array}\right)_{\alpha}
\left( \begin{array}{c}
  c_{p}^{\dagger} \\
  c_{n}^{\dagger} \\
  c_{\bar{p}} \\
  c_{\bar{n}}
  \end{array}\right)_{\alpha} ~,
 \end{equation}
where $u$ and $v$ coefficients are calculated by the following $4\times4$ BCS equation
\begin{equation} \label{eq:hfbeq}
\left( \begin{array}{cccc} \epsilon_{p}-\lambda_{p} & 0 &
\Delta_{p {\bar p}} & \Delta_{p {\bar n}} \\
0  & \epsilon_{n}-\lambda_{n} & \Delta_{n
{\bar p}} & \Delta_{n {\bar n}} \\
  \Delta_{p {\bar p}} &
 \Delta_{p {\bar n}} & -\epsilon_{p} + \lambda_{p} & 0 \\
  \Delta_{n {\bar p}} &
 \Delta_{n {\bar n}} & 0 & -\epsilon_{n} + \lambda_{n}
  \end{array}\right)_{\alpha}
\left( \begin{array}{c}
u_{\alpha'' p} \\ u_{\alpha'' n} \\ v_{\alpha'' p} \\
v_{\alpha'' n} \end{array}\right)_{\alpha}
 =
 E_{\alpha \alpha''}
\left( \begin{array}{c} u_{\alpha'' p} \\ u_{\alpha'' n} \\
 v_{\alpha'' p} \\
v_{\alpha'' n} \end{array}\right)_{\alpha}, 
\end{equation}
where $E_{\alpha \alpha''}$ is an energy of a quasiparticle,  and $\alpha$ indicated by a set of quantum numbers to specify the single-particle-state (SPS).   Since our formalism includes the $np$ pairing correlations, we have two different types of quasiparticles, quasi-proton and quasi-neutron. However the isospins of the quasiparticles cannot be clearly defined \cite{Ha15}. Therefore we denote the isospin indices of quasiparticles as $\alpha'' ( \beta'') = 1,2$ hereafter instead $\tau_Z=\pm1$.   The Greek letters with prime  $(\alpha' , \beta' , \gamma' , \delta')$  are reserved for the isospin of the real particle (see Eqs. (\ref{eq:mat_A}) and (\ref{eq:mat_B})). The pairing potentials in Eq. (\ref{eq:hfbeq}) are permitted between the nucleons in a time-reversed state ($\alpha  {\bar \alpha}$) \cite{Goodmann70}, while the unlike-pairing may have ($\alpha  \alpha$)  pairing as well as (${\bar \alpha}  {\bar \alpha}$) pairing \cite{Goodmann72}, which are effectively included in the present framework. For the mean field we take a Woods-Saxon potential (WS) \cite{cwi} for its usefulness and simplicity. 

The pairing potentials in Eq.(\ref{eq:hfbeq}) are calculated by using the $G$-matrix derived from the realistic Bonn CD potential \cite{Lang93}, which explicitly incorporated the TF by the $\pi$ and $\rho$-meson exchange potentials in the $N$-$N$ interaction,
\begin{equation} \label{eq:gap_BCS}
\Delta_{{p \bar{p}_\alpha}} = \Delta_{\alpha p \bar{\alpha}p} = -
 \Big[ \sum_{J,  c } g_{\textrm{p}} 
G(aacc,J,T=1)\Big] (u_{1p_{c}}^* v_{1p_{c}} +
u_{2p_{c}}^* v_{2p_{c}}) ~,
\end{equation}
\begin{eqnarray} \label{eq:gap_pn_BCS}
\Delta_{{p \bar{n}_\alpha}} = \Delta_{\alpha p \bar{\alpha}n} = &-&
 \Bigg[ \Big[\sum_{J,  c} g_{\textrm{np}}^{T=1} 
G(aacc,J,T=1)\Big] Re(u_{1n_{c}}^* v_{1p_{c}} +
u_{2n_{c}}^* v_{2p_{c}})  \nonumber \\
 &+& \Big[ \sum_{J,  c}
g_{\textrm{np}}^{T=0} iG(aacc,J,T=0)\Big] Im (u_{1n_{c}}^*
v_{1p_{c}} + u_{2n_{c}}^* v_{2p_{c}}) \Bigg]~,
\end{eqnarray}
where the Br\"uckner $G(aacc ~J T)$ matrix in Eqs. (3) and (4) represents the state-dependent pairing matrix elements (PMEs) calculated by using  the spherical basis \cite{Lang93}. $\Delta_{\alpha n \bar{\alpha}n}$ is obtained from Eq. (\ref{eq:gap_BCS}) by replacing $p$ by $n$.

As for the isoscalar $np$ pairing we have two modes, spin-singlet $(S=0)$ and
spin-triplet $(S=1)$. The $S=1$ state comes from the ($\alpha{\alpha}$) and $({\bar
\alpha} {\bar \alpha})$ pairings, which require 8 $\times$ 8
transformation matrix instead of Eq.(3) \cite{Goodmann80}. But, in the present 4 $\times$ 4 scheme, we effectively take into
account the  $T = 0$ channel by the ($\alpha \alpha$) and (${\bar
\alpha} {\bar \alpha}$) channels in the following way. By adopting the procedure in Ref. \cite{Goodmann80} for $np$ and ${\bar n} {\bar
p}$ pairings, we assume
\begin{equation}\label{eq:weight1}
< \alpha n \alpha p, T = 0 | V_{pair} | \beta n \beta p, T = 0 > = < \alpha n \alpha p, T = 0 | V_{pair} | {\bar \beta} n {\bar \beta} p, T = 0 >~.
\end{equation}

Then $\Delta_{\alpha n \alpha p}^{T = 0}$ pairing potential is given as Im $\Delta_{\alpha n \alpha p}^{T = 0} = 0$ and  Re $\Delta_{\alpha n \alpha p}^{T = 0}$ =  Im $\Delta_{\alpha n {\bar \alpha} p}^{T = 0} $ by Eqs. (5) - (7) in Ref. \cite{Goodmann80}. It leads to
\begin{equation}\label{eq:weight2}
\Delta^{2~~ (T= 0)}_{np} = 2| \Delta_{\alpha p {\bar \alpha} n}^{T = 0} |^2 +  2| \Delta_{\alpha p \alpha n}^{T = 0} |^2 = 4 | \Delta_{\alpha p {\bar \alpha} n}^{T = 0} |^2~,
\end{equation}
where a factor two is due to  ${\bar \alpha}p \alpha n$ and ${\bar \alpha}p {\bar \alpha} n$ pairings, respectively.
Therefore we multiply the $g_{np}^{T=0}$ by a factor two, so that  the strength parameters are set as $(g_p, g_n, g_{np}^{T=1}, 2*g_{np}^{T=0} )$, which are  renormalization constants due to the restricted Hilbert model space in practical calculations. Since the characteristic features of the isospin dependence are taken into account already  in the PMEs obtained by the $G$-matrices, we treat the coupling strengths as $g_{np}^{T=0} = g_{np}^{T=1} = g_{np}$. The $g_p$ and $g_n$ are fitted to reproduce the empirical pairing gap, $\Delta_p^{emp}$ and $\Delta_n^{emp}$ evaluated by a five-point mass formula, respectively \cite{Ch93}.

After fixing the $nn$ and $pp$ pairing gaps, theoretical $np$ pairing gaps are calculated as \cite{Ch93,Suhonen98,Bend00}
\begin{equation}
\Delta_{np}^{th.} = - [ ( H_{gs}^{12} + E_1 + E_2 ) - ( H_{gs}^{np} + E_p + E_n)]~.
\label{eq:np}
\end{equation}
Here $H_{gs}^{12} (H_{gs}^{np}) $ is a total BCS ground state energy with (without) $np$ pairing and $ E_1 + E_2 (E_p + E_n)$ is the  sum of the lowest two quasiparticle energies with (without) the $np$ pairing potential $\Delta_{n{\bar p}}$ in Eq.(\ref{eq:hfbeq}). For the complex gap parameters $\Delta_{p {\bar n}_{\alpha}}$ in Eq.(\ref{eq:gap_pn_BCS}), we take their absolute values. Finally the strength parameter $g_{np}$ is fitted to reproduce the empirical $np$ pairing gap data, $\Delta_{np}^{emp}$, which are also deduced from the ground state masses of nearby nuclei. 
We tabulate the $np$ pairing gaps in Table \ref{tab:pairing}, and refer to our previous papers for the detailed calculations of the BCS wave functions including the TF \cite{Ha18-1,Ha18-2}. Hereafter, we briefly summarize the QRPA model, which is applied for calculations of GT strength distributions of the magic nuclei considered in this work.

\begin{table}
\begin{center}
\caption[bb]{Empirical $np$ pairing gaps, $\Delta_{np}^{emp}$, from five-point pairing index formula \cite{Ch93}. The TF effects on the theoretical $np$ pairing gaps, $\Delta_{np}^{th.}$ in Eq.(\ref{eq:np}), obtained by the fitting of $g_{np}$ turn out to be within a few percentage for the empirical pairing gaps.\\}
\setlength{\tabcolsep}{2.0 mm}
\begin{tabular}{ccccc}\hline
  & $^{48}$Ca & $^{90}$Zr & $^{132}$Sn & $^{208}$Pb  \\ \hline \hline
$\Delta_{np}^{emp}$ & 0.418    & 0.190 & 0.409 & 0.237    \\ \hline
 \end{tabular}
\label{tab:pairing}
\end{center}
\end{table}

We start by adopting the standard QRPA formalism based on the equation of motion  of the following phonon operator acting on the BCS ground state \cite{Ha15}
\begin{equation}\label{phonon}
{\cal Q}^{\dagger}_{m,K}  =\sum_{\rho_{\alpha} \alpha \alpha'' \rho_{\beta} \beta \beta''}
[ X^{m}_{( \alpha \alpha'' \beta \beta'')K} A^{\dagger}( \alpha \alpha'' \beta \beta'' K)
- Y^{m}_{( \alpha \alpha'' \beta \beta'')K} {\tilde A}( \alpha \alpha'' \beta \beta'' K)]~,
\end{equation}
where $\rho_{\alpha (\beta)}$ ($\rho_{\alpha (\beta)} = \pm 1$) in the summation denotes the sign of the total angular momentum projection of the given state $\alpha(\beta)$ with pairing creation and annihilation operators composed by two quasiparticles defined as
\begin{equation}
 A^{\dagger}( \alpha \alpha'' \beta \beta'' K)  =
 {[a^{\dagger}_{ \alpha \alpha''} a^{\dagger}_{\beta \beta''}]}^K,
~~~{\tilde A}( \alpha \alpha'' \beta \beta'' K)  =
 {[a_{\beta \beta''} a_{\alpha \alpha''}]}^K,
\end{equation}
where $K$ is the projected quantum number of intrinsic angular momentum on the symmetry axis and a good quantum number in the axially deformed nuclei.

Within the quasi-boson approximation for the phonon operator, we obtain the following QRPA equation including the $np$ pairing correlations in the BCS ground state for describing the correlated QRPA ground state:
\begin{eqnarray}\label{qrpaeq}
&&\left(\begin{array}{cccccccc}
           A_{\alpha \beta \gamma \delta (K)}^{1111} & A_{\alpha \beta \gamma \delta (K)}^{1122} &
           A_{\alpha \beta \gamma \delta (K)}^{1112} & A_{\alpha \beta \gamma \delta (K)}^{1121} &
           B_{\alpha \beta \gamma \delta (K)}^{1111} & B_{\alpha \beta \gamma \delta (K)}^{1122} &
           B_{\alpha \beta \gamma \delta (K)}^{1112} & B_{\alpha \beta \gamma \delta (K)}^{1121} \\
           A_{\alpha \beta \gamma \delta (K)}^{2211} & A_{\alpha \beta \gamma \delta (K)}^{2222} &
           A_{\alpha \beta \gamma \delta (K)}^{2212} & A_{\alpha \beta \gamma \delta (K)}^{2221} &
           B_{\alpha \beta \gamma \delta (K)}^{2211} & B_{\alpha \beta \gamma \delta (K)}^{2222} &
           B_{\alpha \beta \gamma \delta (K)}^{2212} & B_{\alpha \beta \gamma \delta (K)}^{2221}\\
           A_{\alpha \beta \gamma \delta (K)}^{1211} & A_{\alpha \beta \gamma \delta (K)}^{1222} &
           A_{\alpha \beta \gamma \delta (K)}^{1212} & A_{\alpha \beta \gamma \delta (K)}^{1221} &
           B_{\alpha \beta \gamma \delta (K)}^{1211} & B_{\alpha \beta \gamma \delta (K)}^{1222} &
           B_{\alpha \beta \gamma \delta (K)}^{1212} & B_{\alpha \beta \gamma \delta (K)}^{1221}\\
           A_{\alpha \beta \gamma \delta (K)}^{2111} & A_{\alpha \beta \gamma \delta (K)}^{2122} &
           A_{\alpha \beta \gamma \delta (K)}^{2112} & A_{\alpha \beta \gamma \delta (K)}^{2121} &
           B_{\alpha \beta \gamma \delta (K)}^{2111} & B_{\alpha \beta \gamma \delta (K)}^{2122} &
           B_{\alpha \beta \gamma \delta (K)}^{2112} & B_{\alpha \beta \gamma \delta (K)}^{2121} \\
             &       &       &      &      &        &           &        \\ \nonumber
           - B_{\alpha \beta \gamma \delta (K)}^{1111} & -B_{\alpha \beta \gamma \delta (K)}^{1122} &
            -B_{\alpha \beta \gamma \delta (K)}^{1112} & -B_{\alpha \beta \gamma \delta (K)}^{1121} &
           - A_{\alpha \beta \gamma \delta (K)}^{1111} & -A_{\alpha \beta \gamma \delta (K)}^{1122} &
           -A_{\alpha \beta \gamma \delta (K)}^{1112}  & -A_{\alpha \beta \gamma \delta (K)}^{1121}\\
           - B_{\alpha \beta \gamma \delta (K)}^{2211} & -B_{\alpha \beta \gamma \delta (K)}^{2222} &
           -B_{\alpha \beta \gamma \delta (K)}^{2212}  & -B_{\alpha \beta \gamma \delta (K)}^{2221} &
           - A_{\alpha \beta \gamma \delta (K)}^{2211} & -A_{\alpha \beta \gamma \delta (K)}^{2222} &
           -A_{\alpha \beta \gamma \delta (K)}^{2212}  & -A_{\alpha \beta \gamma \delta (K)}^{2221}\\
           - B_{\alpha \beta \gamma \delta (K)}^{1211} & -B_{\alpha \beta \gamma \delta (K)}^{1222} &
           -B_{\alpha \beta \gamma \delta (K)}^{1212}  & -B_{\alpha \beta \gamma \delta (K)}^{1221} &
           - A_{\alpha \beta \gamma \delta (K)}^{1211} & -A_{\alpha \beta \gamma \delta (K)}^{1222} &
           -A_{\alpha \beta \gamma \delta (K)}^{1212}  & -A_{\alpha \beta \gamma \delta (K)}^{1221} \\
          - B_{\alpha \beta \gamma \delta (K)}^{2111} & -B_{\alpha \beta \gamma \delta (K)}^{2122} &
           -B_{\alpha \beta \gamma \delta (K)}^{2112}  & -B_{\alpha \beta \gamma \delta (K)}^{2121} &
           - A_{\alpha \beta \gamma \delta (K)}^{2111} & -A_{\alpha \beta \gamma \delta (K)}^{2122} &
           -A_{\alpha \beta \gamma \delta (K)}^{2112}  & -A_{\alpha \beta \gamma \delta (K)}^{2121} \\
           \end{array} \right)\\  && \times
\left( \begin{array}{c}   {\tilde X}_{(\gamma 1 \delta 1)K}^{m}  \\ {\tilde X}_{(\gamma 2 \delta 2)K}^{m} \\
  {\tilde X}_{(\gamma 1 \delta 2)K}^{m} \\  {\tilde X}_{(\gamma 2 \delta 1)K}^{m} \\ \\
     {\tilde Y}_{(\gamma 1 \delta 1)K}^{m} \\ {\tilde Y}_{(\gamma 2 \delta 2)K}^{m} \\
     {\tilde Y}_{(\gamma 1 \delta 2)K}^{m}\\{\tilde Y}_{(\gamma 2 \delta 1)K}^{m} \end{array} \right)
 = \hbar {\Omega}_K^{m}
 \left ( \begin{array}{c} {\tilde X}_{(\alpha 1 \beta 1)K}^{m}  \\{\tilde X}_{(\alpha 2 \beta 2)K}^{m} \\
 {\tilde X}_{(\alpha 1 \beta 2)K}^{m} \\  {\tilde X}_{(\alpha 2 \beta 1)K}^{m}\\ \\
{\tilde Y}_{(\alpha 1 \beta 1)K}^{m} \\ {\tilde Y}_{(\alpha 2 \beta 2)K}^{m} \\
{\tilde Y}_{(\alpha 1 \beta 2)K}^{m} \\ {\tilde Y}_{(\alpha 2 \beta 1)K}^{m} \end{array} \right)  ~,
\end{eqnarray}
where the amplitudes
${\tilde X}^m_{(\alpha \alpha''  \beta \beta'')K }$ and ${\tilde Y}^m_{(\alpha
\alpha''  \beta \beta'')K}$ in Eq. (\ref{qrpaeq}) stand for the forward and backward going amplitudes, respectively,  from a state ${ \alpha
\alpha'' }$ to another state ${\beta  \beta''}$ \cite{Ha15}. The forward and backward amplitudes $X^m_{(\alpha \alpha''
 \beta \beta'')K}$ and $Y^m_{(\alpha \alpha''
 \beta \beta'')K}$ in the phonon operator in Eq. (\ref{phonon}) are related to
$\tilde{X^m}_{(\alpha \alpha'' \beta \beta'')K}=\sqrt2 \sigma_{\alpha \alpha'' \beta \beta''} X^m_{(\alpha \alpha''
 \beta \beta'')K}$
and $\tilde{Y^m}_{(\alpha \alpha'' \beta \beta'')K}=\sqrt2 \sigma_{\alpha \alpha'' \beta \beta''}
Y^m_{(\alpha \alpha'' \beta \beta'')K}$ in Eq. (\ref{qrpaeq}), where $\sigma_{\alpha \alpha'' \beta \beta''}$ = 1 if $\alpha = \beta$ and $\alpha''$ =
$\beta''$, otherwise $\sigma_{\alpha \alpha'' \beta \beta'' }$ = $\sqrt 2$ \cite{Ha15}.

If we switch off the $np$ pairing, all off-diagonal terms in the A and B matrices in Eq. (\ref{qrpaeq}) disappear with the replacement of 1 and 2 into proton and neutron. Then the QRPA equation is decoupled into $\bf{pp + nn + pn + np}$ QRPA equations. The ${\bf pp + nn}$ QRPA can describe charge conserving reactions such as the M1 spin or EM transitions on the same nuclear species, while ${\bf np + pn}$ QRPA describes charge exchange reactions like the GT${(+/-)}$  transitions. The A and B matrices in Eq. (\ref{qrpaeq}) are given by
\begin{eqnarray} \label{eq:mat_A}
A_{\alpha \beta \gamma \delta (K)}^{\alpha'' \beta'' \gamma'' \delta''}  = && (E_{\alpha
   \alpha''} + E_{\beta \beta''}) \delta_{\alpha \gamma} \delta_{\alpha'' \gamma''}
   \delta_{\beta \delta} \delta_{\beta'' \delta''}
       - \sigma_{\alpha \alpha'' \beta \beta''}\sigma_{\gamma \gamma'' \delta \delta''}\\ \nonumber
   &\times&
   \sum_{\alpha' \beta' \gamma' \delta'}
   [-g_{pp} (u_{\alpha \alpha''\alpha'} u_{\beta \beta''\beta'} u_{\gamma \gamma''\gamma'} u_{\delta \delta''\delta'}
   +v_{\alpha \alpha''\alpha'} v_{\beta \beta''\beta'} v_{\gamma \gamma''\gamma'} v_{\delta \delta''\delta'} )
    ~V_{\alpha \alpha' \beta \beta',~\gamma \gamma' \delta \delta'}
    \\ \nonumber  &-& g_{ph} (u_{\alpha \alpha''\alpha'} v_{\beta \beta''\beta'}u_{\gamma \gamma''\gamma'}
     v_{\delta \delta''\delta'}
    +v_{\alpha \alpha''\alpha'} u_{\beta \beta''\beta'}v_{\gamma \gamma''\gamma'} u_{\delta \delta''\delta'})
    ~V_{\alpha \alpha' \delta \delta',~\gamma \gamma' \beta \beta'}
     \\ \nonumber  &-& g_{ph} (u_{\alpha \alpha''\alpha'} v_{\beta \beta''\beta'}v_{\gamma \gamma''\gamma'}
     u_{\delta \delta''\delta'}
     +v_{\alpha \alpha''\alpha'} u_{\beta \beta''\beta'}u_{\gamma \gamma''\gamma'} v_{\delta \delta''\delta'})
    ~V_{\alpha \alpha' \gamma \gamma',~\delta \delta' \beta \beta' }],
\end{eqnarray}
\begin{eqnarray} \label{eq:mat_B}
B_{\alpha \beta \gamma \delta (K)}^{\alpha'' \beta'' \gamma'' \delta''}  =
 &-& \sigma_{\alpha \alpha'' \beta \beta''} \sigma_{\gamma \gamma'' \delta \delta''}
  \\ \nonumber &\times&
 \sum_{\alpha' \beta' \gamma' \delta'}
  [g_{pp}
  (u_{\alpha \alpha''\alpha'} u_{\beta \beta''\beta'}v_{\gamma \gamma''\gamma'} v_{\delta \delta''\delta'}
   +v_{\alpha \alpha''\alpha'} v_{{\bar\beta} \beta''\beta'}u_{\gamma \gamma''\gamma'} u_{{\bar\delta} \delta''\delta'} )
   ~ V_{\alpha \alpha' \beta \beta',~\gamma \gamma' \delta \delta'}\\ \nonumber
     &- & g_{ph} (u_{\alpha \alpha''\alpha'} v_{\beta \beta''\beta'}v_{\gamma \gamma''\gamma'}
     u_{\delta \delta''\delta'}
    +v_{\alpha \alpha''\alpha'} u_{\beta \beta''\beta'}u_{\gamma \gamma''\gamma'} v_{\delta \delta''\delta'})
   ~ V_{\alpha \alpha' \delta \delta',~\gamma \gamma' \beta \beta'}
     \\ \nonumber  &- & g_{ph} (u_{\alpha \alpha''\alpha'} v_{\beta \beta''\beta'}u_{\gamma \gamma''\gamma'}
      v_{\delta \delta''\delta'}
     +v_{\alpha \alpha''\alpha'} u_{\beta \beta''\beta'}v_{\gamma \gamma''\gamma'} u_{\delta \delta''\delta'})
   ~ V_{\alpha \alpha' \gamma \gamma',~\delta \delta' \beta \beta'}],
\end{eqnarray}
where $u$ and $v$ coefficients are determined from BCS Equation. The $g_{pp}$ and $g_{ph}$ stand for $p-p$ and $p-h$ renormalization factors for the residual interactions in Eqs. (\ref{eq:mat_A}) and (\ref{eq:mat_B}).  The two-body interactions $V_{\alpha \beta,~\gamma \delta}$ and $V_{\alpha \delta,~\gamma \beta}$ are $p-p$ and $p-h$ matrix elements of the residual $N$-$N$ interaction $V$, respectively, which are calculated from the $G$-matrix.

The $G$-matrix is obtained by using the Br\"uckner reaction matrix based on the realistic Bonn potential for the $N$-$N$ interaction inside nuclei from the Bethe-Goldstone equation as follows,
\begin{equation}
G (W)_{\alpha \beta,~\gamma \delta} = V_{\alpha \beta,~\gamma \delta}^{OBEP} + V_{\alpha \beta,~\gamma \delta}^{OBEP} {  {Q_p } \over {W- H_0 -i \epsilon}} G (W)_{\alpha \beta,~\gamma \delta},
\end{equation}
where $\alpha, \beta, \gamma, \delta$ indicate the associated quantum numbers of single-nucleon basis states as stated before (oscillator wave functions
with single-particle energies from a Woods-Saxon potential). $H_0$ is the harmonic oscillator Hamiltonian and $Q_p$ is the Pauli operator. For finite nuclei, the starting
energy $W$ is chosen as an average single-particle energy. $V_{\alpha \beta,~\gamma \delta}^{OBEP}$ is the one
boson exchange potential of the Bonn group \cite{Lang93}. The repulsive (attractive) TF effects between $j_{>(<)}$ and $j_{>(<)}$ ($j_{>(<)}$ and $j_{<(>)}$) states appear explicitly on the $G$-matrix as shown in Fig.4 of Ref. \cite{Ha21}.

The GT transition amplitudes from the ground state of an initial (parent) nucleus
to the excited state of a daughter nucleus, {\it i.e.},  the one phonon state
$K^+$ in a daughter nucleus, are written as \cite{Ha15}
\begin{eqnarray} \label{eq:phonon}
&&< K^+, m | {\hat {GT}}_{K }^- | ~QRPA >  \\ \nonumber
&&= \sum_{\alpha \alpha''\rho_{\alpha} \beta \beta''\rho_{\beta}}{\cal N}_{\alpha \alpha''\rho_{\alpha}
 \beta \beta''\rho_{\beta} }
 < \alpha \alpha''p \rho_{\alpha}|  \sigma_K | \beta \beta''n \rho_{\beta}>
 [ u_{\alpha \alpha'' p} v_{\beta \beta'' n} X_{(\alpha \alpha''\beta \beta'')K}^{m} +
v_{\alpha \alpha'' p} u_{\beta \beta'' n} Y_{(\alpha \alpha'' \beta \beta'')K}^{m}], \\ \nonumber
&&< K^+, m | {\hat {GT}}_{K }^+ | ~QRPA >  \\ \nonumber
&&= \sum_{\alpha \alpha'' \rho_{\alpha} \beta \beta''\rho_{\beta}}{\cal N}_{\alpha \alpha'' \beta \beta'' }
 < \alpha \alpha''p \rho_{\alpha}|  \sigma_K | \beta \beta''n \rho_{\beta}>
 [ u_{\alpha \alpha'' p} v_{\beta \beta'' n} Y_{(\alpha \alpha'' \beta \beta'')K}^{m} +
v_{\alpha \alpha'' p} u_{\beta \beta'' n} X_{(\alpha \alpha'' \beta \beta'')K}^{m} ]~,
\end{eqnarray}
where %$K^{\mp}$ is 
 the $|~QRPA >$ denotes the correlated QRPA ground state in an intrinsic frame and
the normalization factor is given as $ {\cal N}_{\alpha \alpha'' \beta
 \beta''} (J,T) = \sqrt{ 1 - \delta_{\alpha \beta} \delta_{\alpha'' \beta''} (-1)^{J + T} }/
 (1 + \delta_{\alpha \beta} \delta_{\alpha'' \beta''})$ with total isospin $T$. The particle model space was extended up to  $N_0 = 10 \hbar \omega$ in the spherical basis.
%

%%%%%%%%%%%%%%%%%%%%%%%%%%%%%%%%%%%%%%%%%%%%%%%%%%

\section{Results}
Before discussing numerical results concerning the TF effect, we briefly explain how to separate the TF effect in the $G$-matrix calculation based on the CD Bonn potential \cite{Mach87,Lang93}. Because the TF does not accommodate well  in the $j$-$j$ coupling scheme \cite{Alex85}, we exploit the $N$-$N$ potential  represented by the $L$-$S$ coupling scheme, $|LSJM>$ basis \cite{Mach87}. The $N$-$N$ potential, which is a summation of the contributions of  meson-exchange potentials given by the six helicity amplitudes, is decomposed by spin-singlet, uncoupled spin-triplet and coupled spin-triplet channels \cite{Lang93}. Here the uncoupled and the coupled spin-triplet channels correspond to the cases of $J=L$ and $J=L \pm 1$, respectively. The coupled case mainly comes from the $\rho$- and $\pi$-exchange potentials and corresponds to the TF  comprising $\Delta L =2$ and $\Delta L= 0$ components \cite{Lang93}.
Inside nuclei, not only $J=1$,  but also other $J$-coupling TFs play significant roles.

\begin{figure}
\centering
\includegraphics[width=0.45\linewidth]{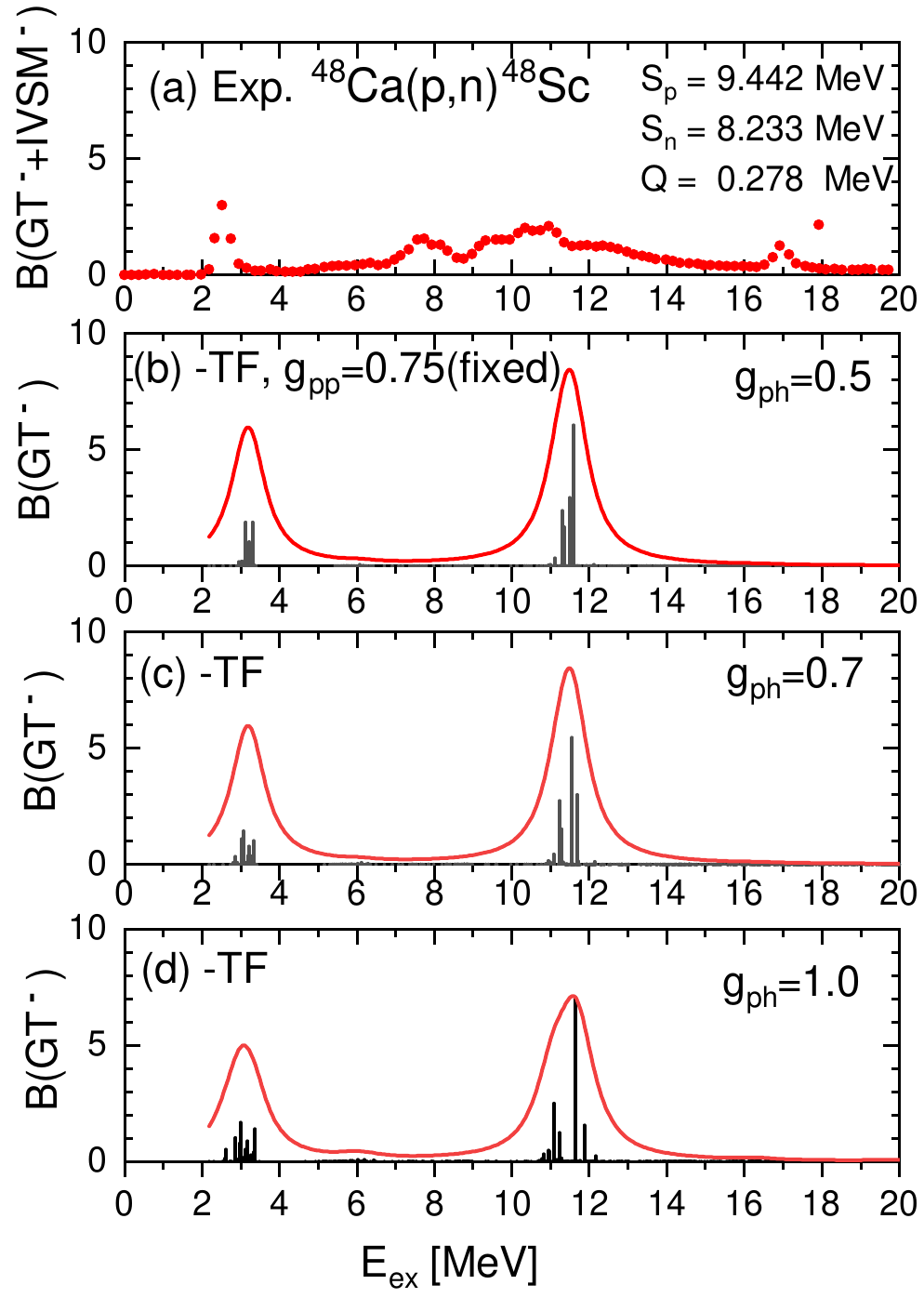}
\includegraphics[width=0.45\linewidth]{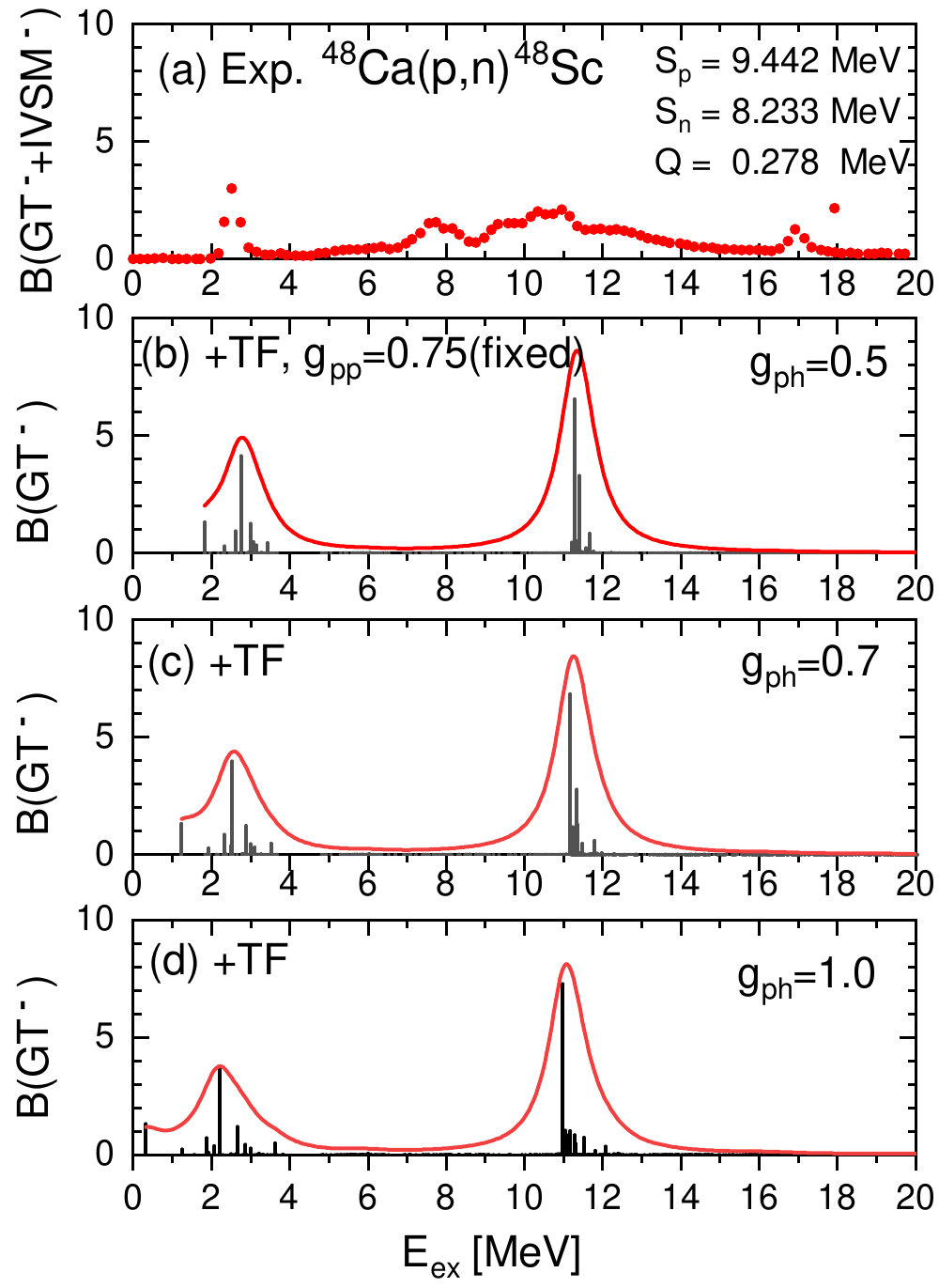}
\caption{(Color online) GT(-) transition strength distribution B(GT(-)) of $^{48}$Ca. Experimental data by (p,n) reaction in panel (a) are taken from Ref. \cite{Yako09}.
Results of (b), (c), and (d) are without (left panel) and with (right panel) the TF. The normalization factor $g_{pp}$ is fixed in all the calculations,  but the factor $g_{ph}$ is changed as 0.5, 0.7 and  1.0
for the panels (b), (c) and  (d), respectively. The calculated results are smoothed out by a Lorentzian function with the width parameter of 1 MeV and shown by a red curve. The experimental data contain the isovector spin-monopole strength (IVSM) in the forward angle cross sections. However the excitation energy of dominant IVSM peak is expected at around 35 MeV in $^{48}$Ca(p,n)$^{48}$Sc reaction \cite{Hama2000}, and the IVSM has very small  strength below 20 MeV, and does not affect much on  the present discussion.} \label{fig1}
\end{figure}

\begin{figure}
\centering
\includegraphics[width=0.49\linewidth]{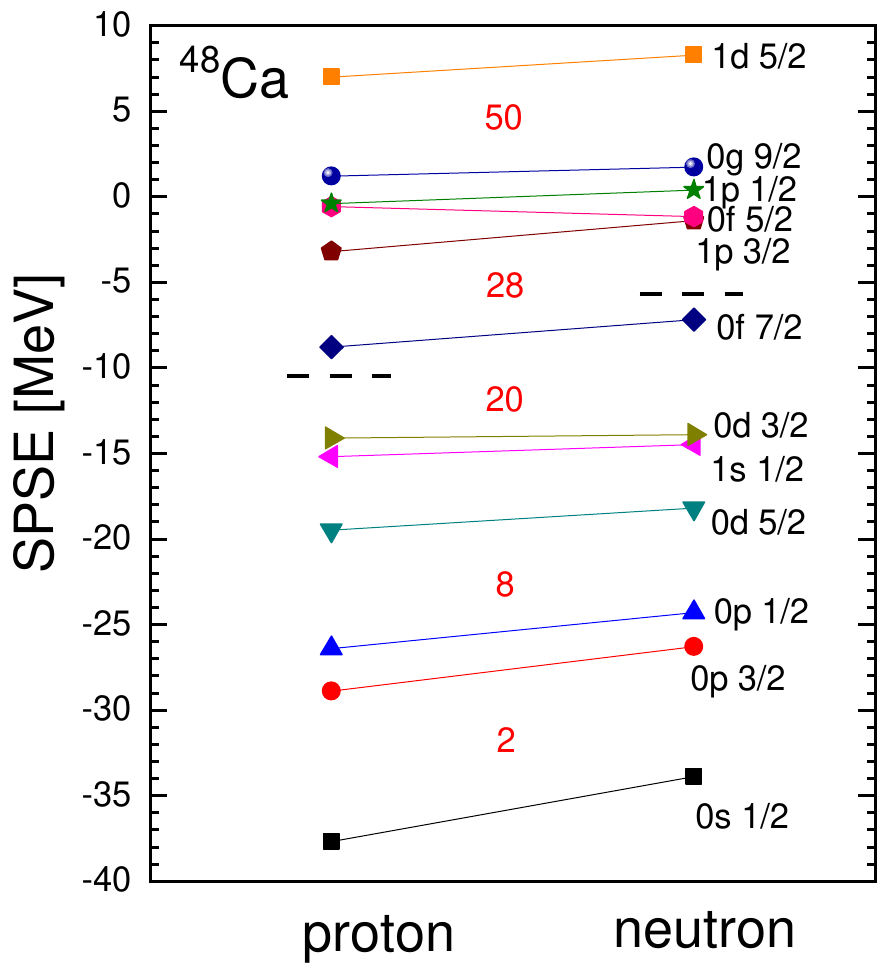}\includegraphics[width=0.49\linewidth]{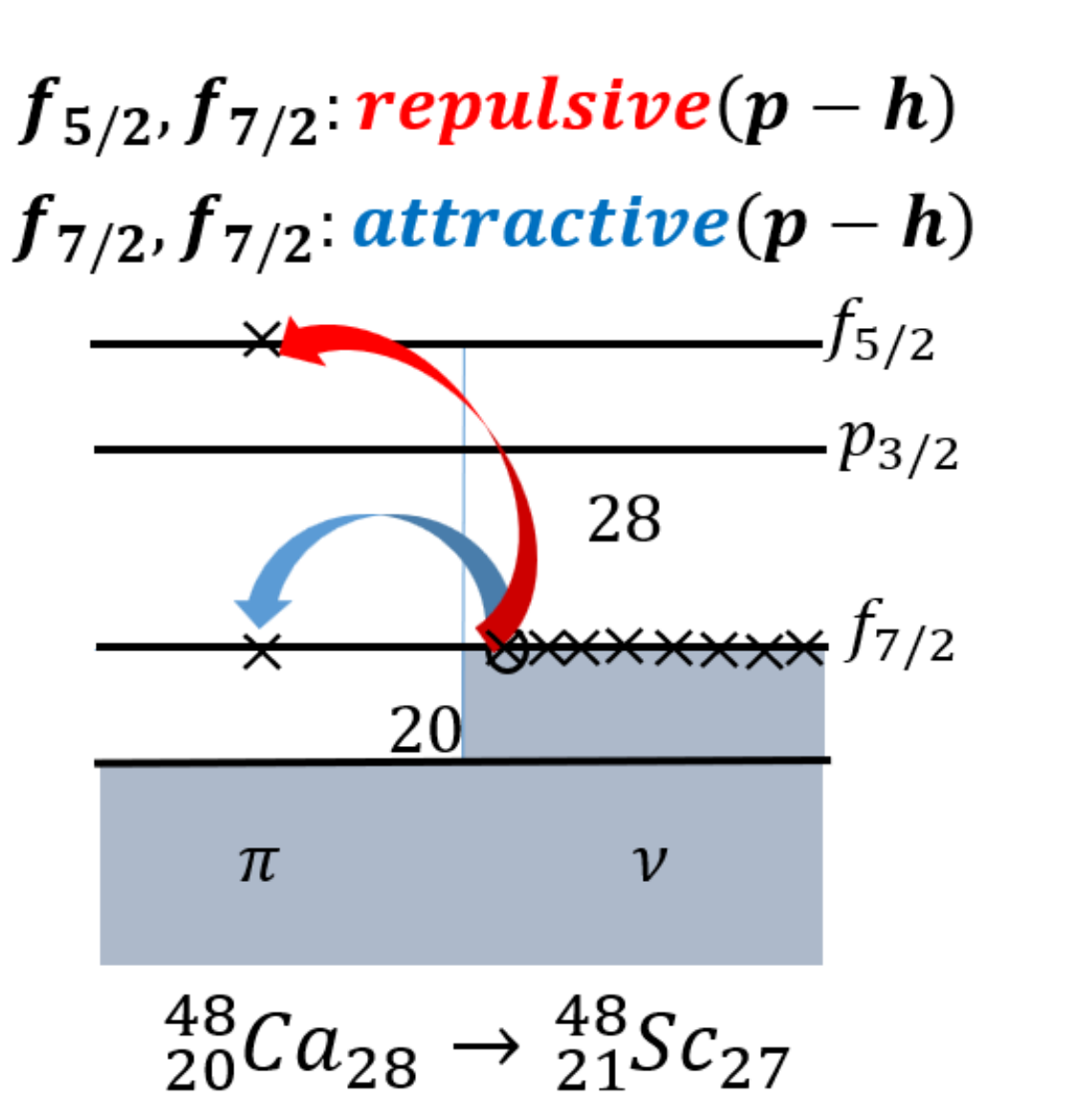}
\caption{(Color online)  (left) Calculated SPS energies by using the WS potential \cite{cwi}, where the black dashed lines indicate with Fermi energies. (right) Schematic diagram right of GT transition for $^{48}$Ca. The blue and red arrows in the right panel represent the main $p-h$ configurations of GT transition, respectively, at $E_{\textrm{ex}}$=2.5 and 10.5 MeV. For magic nuclei, the repulsive (attractive) $p-p$ TF between $j^{\nu}_{>(<)}$ and  $j^{\pi}_{>(<)}$ ($j^{\nu}_{>(<)}$ and  $j^{\pi}_{<(>)}$) states is changed to the attractive (repulsive) $p-h$ TF by the Pandya transformation.}
\label{fig2}
\end{figure}

\subsection{GT strength distribution of $^{48}$Ca}

\begin{figure}
\centering
\includegraphics[width=0.45\linewidth]{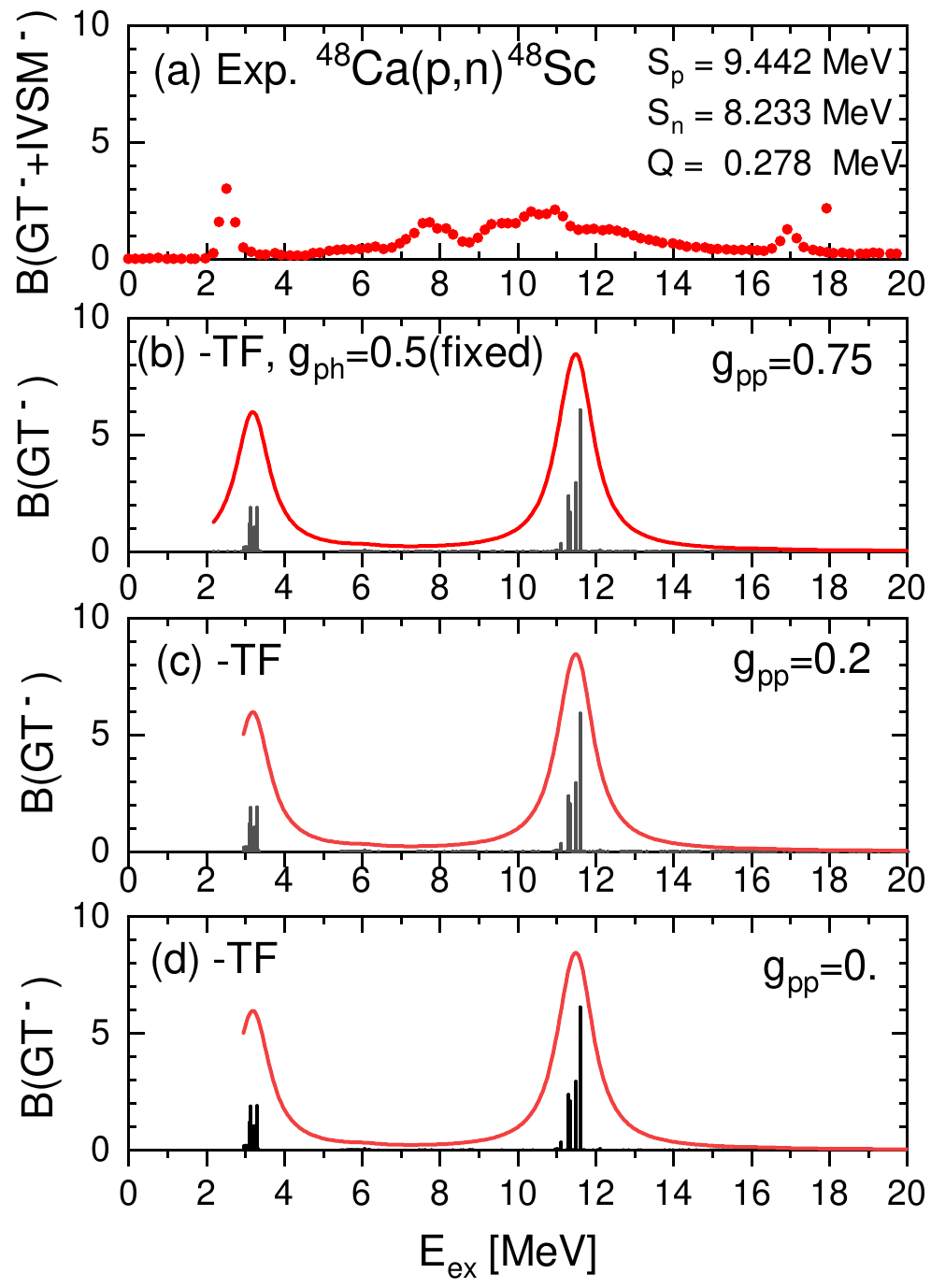}
\includegraphics[width=0.45\linewidth]{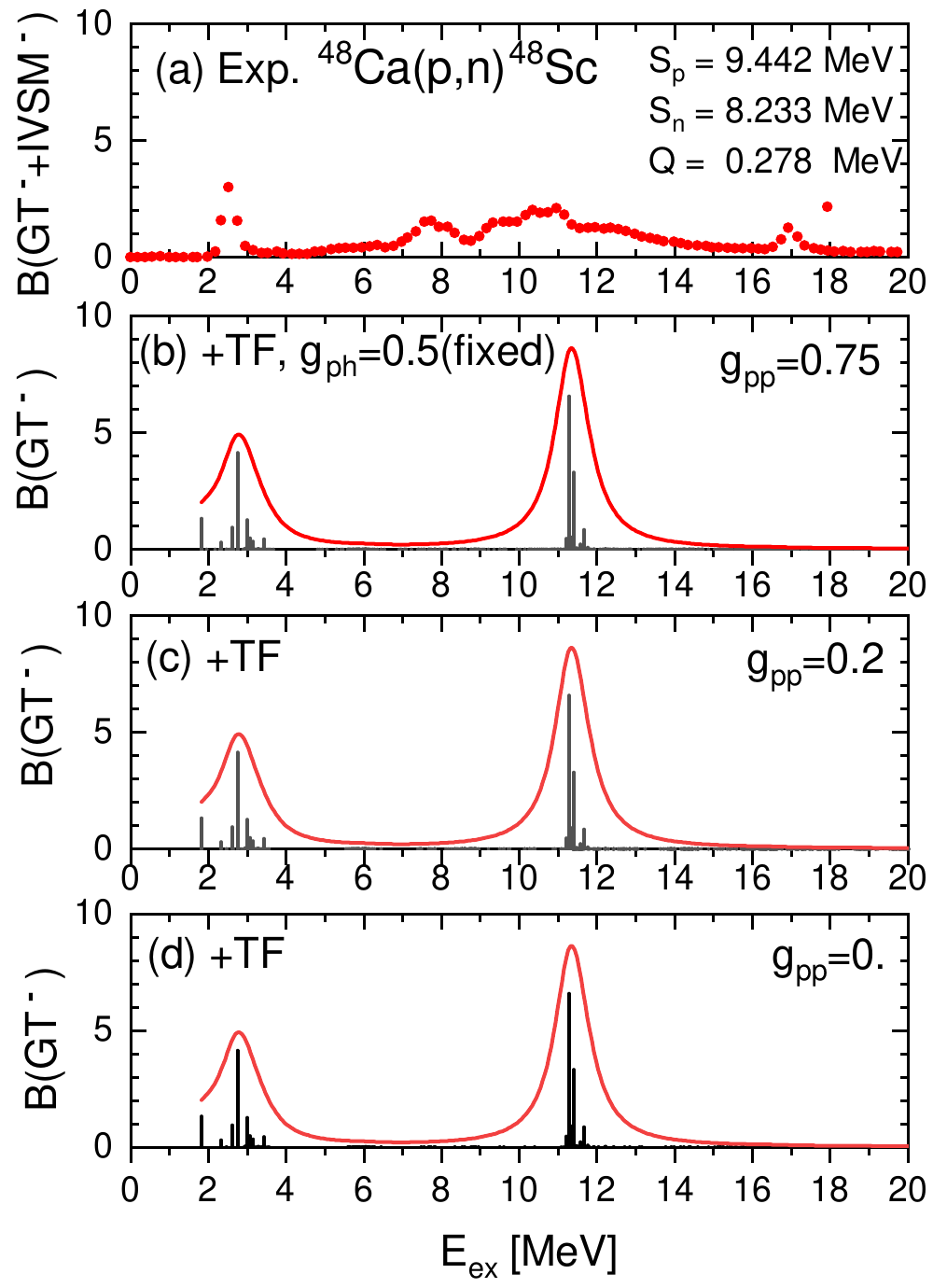}
\caption{(Color online) Effect of the $p-p$ interaction on the GT(-) transition strength distribution B(GT(-)) of $^{48}$Ca. Left (right) panels are the results w/o TF and with TF. Results of (b), (c), and (d) show the results where the normalization factor $g_{ph}$= 1.0 is fixed, but the factor $g_{pp}$ is changed as 0., 0.2 and 0.75.  Others are the same as those in Fig.\ref{fig1}.} \label{fig3}
\end{figure}

In Fig.\ref{fig1}, we present the TF effect on the GT strength distribution for $^{48}$Ca including the TF in the residual interaction by the $G$-matrix.
The GT transition operator is defined by
\be
\hat{O}({GT}_{\pm})=\sum_i {\bf \sigma t}_{\pm},
\ee
for the charge exchange $(n,p)$ and $(p,n)$ channels, respectively.
We notice that $^{48}$Ca is a doubly-magic nucleus where the $p-h$ interaction becomes significant for GT transition. Therefore  while the $p-p $ strength is fixed as $g_{pp}$ = 0.75,  the $p-h$ strength, $g_{ph}$, is varied  from 0.5 to 1.0. These variations are multiplied by the QRPA matrix elements for the $p-p$ and $p-h$ interactions in the QRPA equation, respectively, in Eqs. (\ref{eq:mat_A}) and (\ref{eq:mat_B}).

Experimental GT strength distribution data in Fig.\ref{fig1} (a) shows a small peak in low-lying energy region around $E_{\textrm{ex}}=2.5$ MeV,  and also main
broad peak at $E_{\textrm{ex}}= 7-15$ MeV.   Main configuration for the low-energy peak is the $p-h$ excitation, $\nu 0 f_{7/2} \rightarrow \pi 0 f_{7/2}$, and the high-energy peak is
dominated by the spin-flip $p-h$ excitation, $\nu 0 f_{7/2} \rightarrow \pi 0 f_{5/2}$.  We study the role of TF on these GT peaks.  In the left panel of Fig.\ref{fig1}, the $p-h$ interaction is changed by a factor $g_{ph}$=0.5, 0.7 and 1.0 for Fig.\ref{fig1}(b), (c) and (d), respectively without TF.
Without the TF, the positions of the low-energy and high-energy peaks are  not so much changed even with the increase of the $g_{ph}$ strength. With the TF, the low-energy peak is slightly enriched and more fragmented,
   and with the increase of the $g_{ph}$ strength the low-lying state is slightly shifted to lower energy region as shown in the right panel of Fig.\ref{fig1}. This shift is attributed to the attractive $p-h$ TF property on the same $j_\pi =j_\nu$ as displayed in Fig.\ref{fig2}. 
   
High-lying GT states are not so much affected  by the TF.  The fragments come from the $np$ pairing containing the tensor force in the residual
interaction. The configuration of the GT states is increased because of slight melting of Fermi surfaces of protons and neutrons by the $np$ pairing correlations in HFB calculations, which gives
rise to the fragments, although the TF turns out to affect only a few percentage in the $np$ pairing gap. For reference we tabulated the $np$ pairing gaps for the nuclei exploited in this work in Table \ref{tab:pairing}. Moreover the off-diagonal matrix in the QRPA equation may enhance such fragmentations.

The GT transition configurations for the low-lying GT state at $E_{\textrm{ex}}=2.5$ MeV for $^{48}$Ca is represented as the blue arrow in Fig.\ref{fig2}, whose transition is on the same shell orbit ($\nu 0 f_{7/2} \rightarrow \pi 0 f_{7/2}$), and TF makes
the attractive $p-h$ residual interaction. The high-lying state is dominated by  the transition of the SO partners ($\nu 0 f_{7/2} \rightarrow \pi 0 f_{5/2}$) with the repulsive ${p-h}$ residual interaction (see red arrow).

In Fig.\ref{fig3}, we show that the GT strength distributions are almost insensitive to the change of $g_{pp}$ strength.
It means that the repulsive $p-p$ TF on the same $f_{7/2}$ shell is not active because of the closed shell structure of $^{48}$Ca. Instead, the TF appears as the attractive $p-h$ TF as illustrated in Fig.\ref{fig2}. This feature comes from the following Pandya transformation between $p-h$ and $p-p$  matrix elements,
\begin{eqnarray}
< \alpha \beta^{-1} ; JT | V_{res.} | \gamma \delta^{-1} ; JT  >
&=& - \Sigma_{J' T'} {\tilde {J'}}^2 {\tilde {T'}}^2 W(j_{\alpha} j_{\beta} j_{\gamma} j_{\delta}; J J^{'})
W ( {1 \over 2}{1 \over 2}{1 \over 2}{1 \over 2} ; T T^{'}) \\ \nonumber
& & \times < \alpha \delta ; J^{'} T^{'} | V_{res.} | \beta \gamma ; J^{'} T^{'} >~.
\end{eqnarray}

%%%%%%%%%%%%%%%%%%%%%%%%%%%%%%%%%%%%%%%%%%%%

\begin{figure}
\centering
\includegraphics[width=0.45\linewidth]{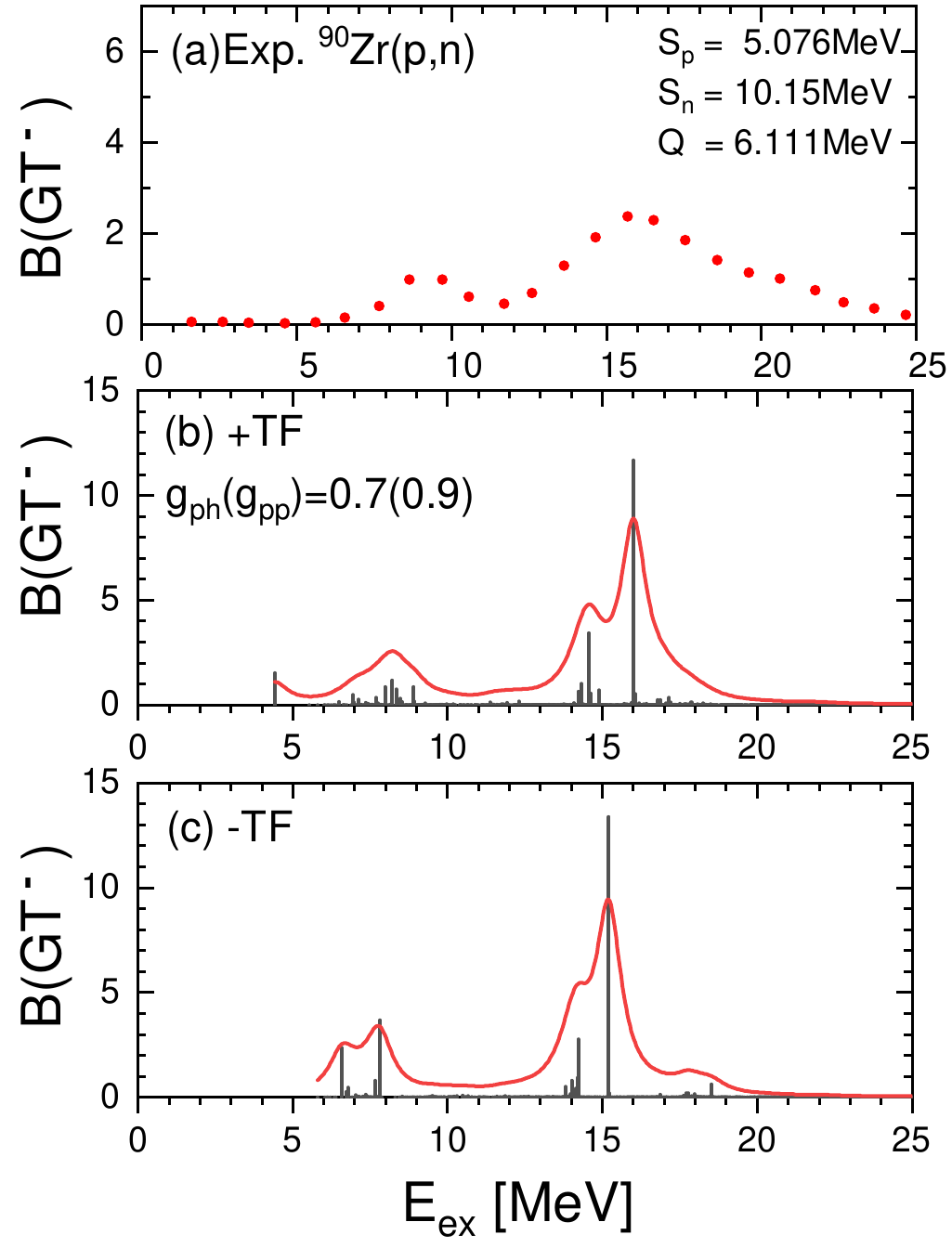}
\includegraphics[width=0.45\linewidth]{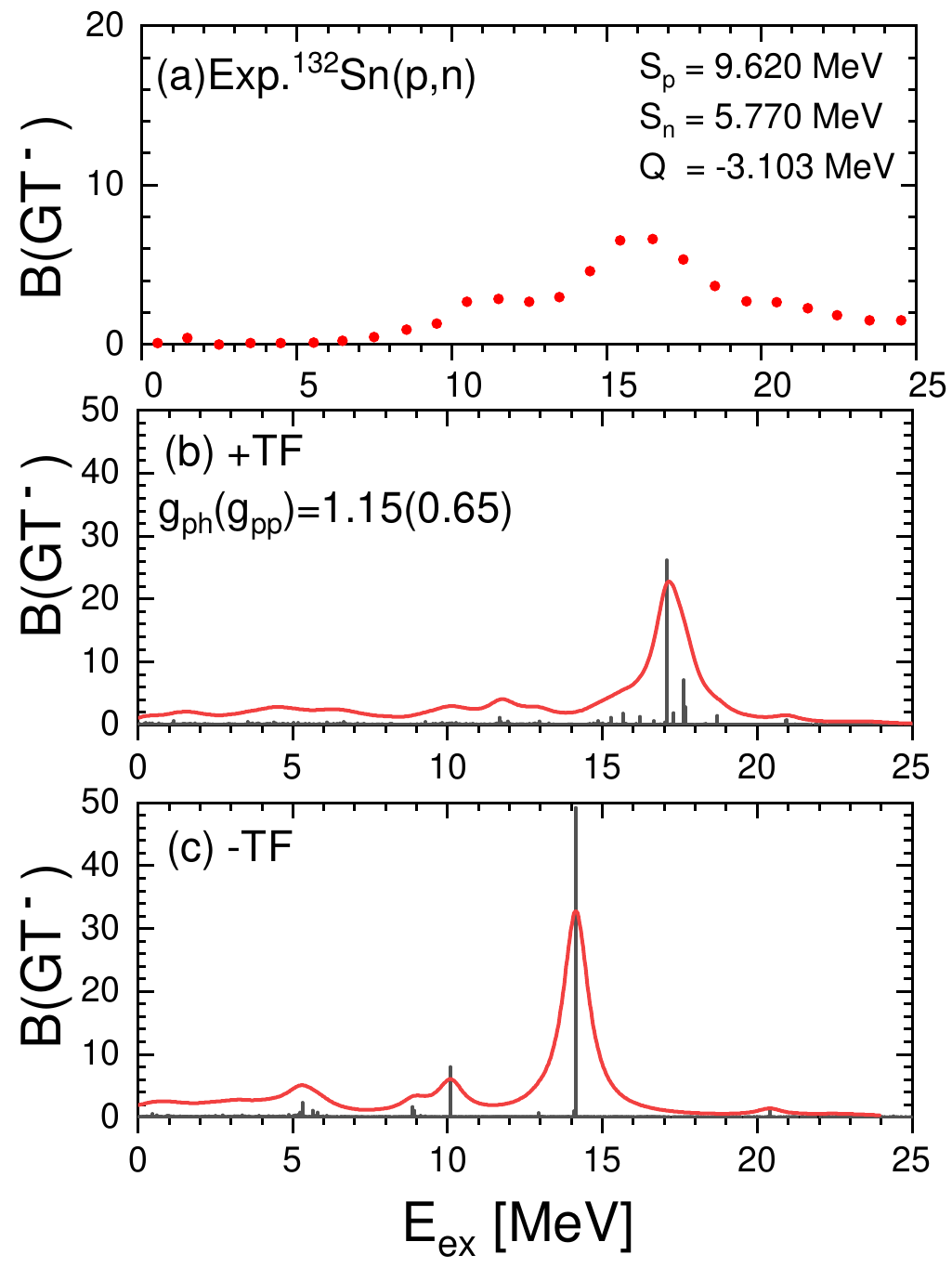}
\caption{(Color online) The TF effects on the GT strength distributions for $^{90}$Zr (left panel) and $^{132}$Sn (right panel). The panel (b) and (c) correspond to  the cases with and without the TF, respectively.   The normalization factors are fixed to be $g_{ph}=$ 0.7(1.15) and  $g_{pp}=$ 0.9 (0.65) for $^{90}$Zr ($^{132}$Sn), respectively.
Experimental data by (p,n) reaction are from Ref. \cite{Wakasa97} and Ref. \cite{Yasuda18}, respectively.}
\label{fig4}
\end{figure}

\subsection{GT strength distribution of $^{90}$Zr, $^{132}$Sn, and $^{208}$Pb}

% for 90Zr and $^{132}$Sn case
In the left panel of Fig.\ref{fig4}(a) for $^{90}$Zr, there appears a left shoulder of the main GT state around $E_{\textrm{ex}}=$ 9 in the experimental data.  If we include the TF in the RPA calculation, the left shoulder shifts slightly to the low-energy GT region in Fig.\ref{fig4}(b) compared with Fig.\ref{fig4}(c).  We notice that  the shoulder around $E_{\textrm{ex}}=9$ MeV is fragmented by  the TF.
For $^{90}$Zr case, the low-lying state is the transition from $\nu 0g_{9/2}$ to $\pi 0g_{9/2}$,  and high-lying state comes from $\nu 0g_{9/2}$ to $\pi 0g_{7/2}$ state, which is similar to the case of $^{48}$Ca. %But the shift is larger than that of $^{48}$Ca.
The high-lying GT peak is shifted about 1 MeV higher in energy   by  the repulsive $p-h$ TF and becomes close %by which the GT peak is shifted to the higher states near
to the experimental peak position.
We found almost the same mechanism also for $^{132}$Sn case.  The low-energy strength is more fragmented by the TF and the main GT peak is shifted to higher energy by about 3 MeV and becomes consistent with the experimental data.  The TF effect is more enhanced in the heavier nuclei as shown in the shift of the main GT peaks in Fig.\ref{fig4}.

\begin{figure}
\centering
\includegraphics[width=0.65\linewidth]{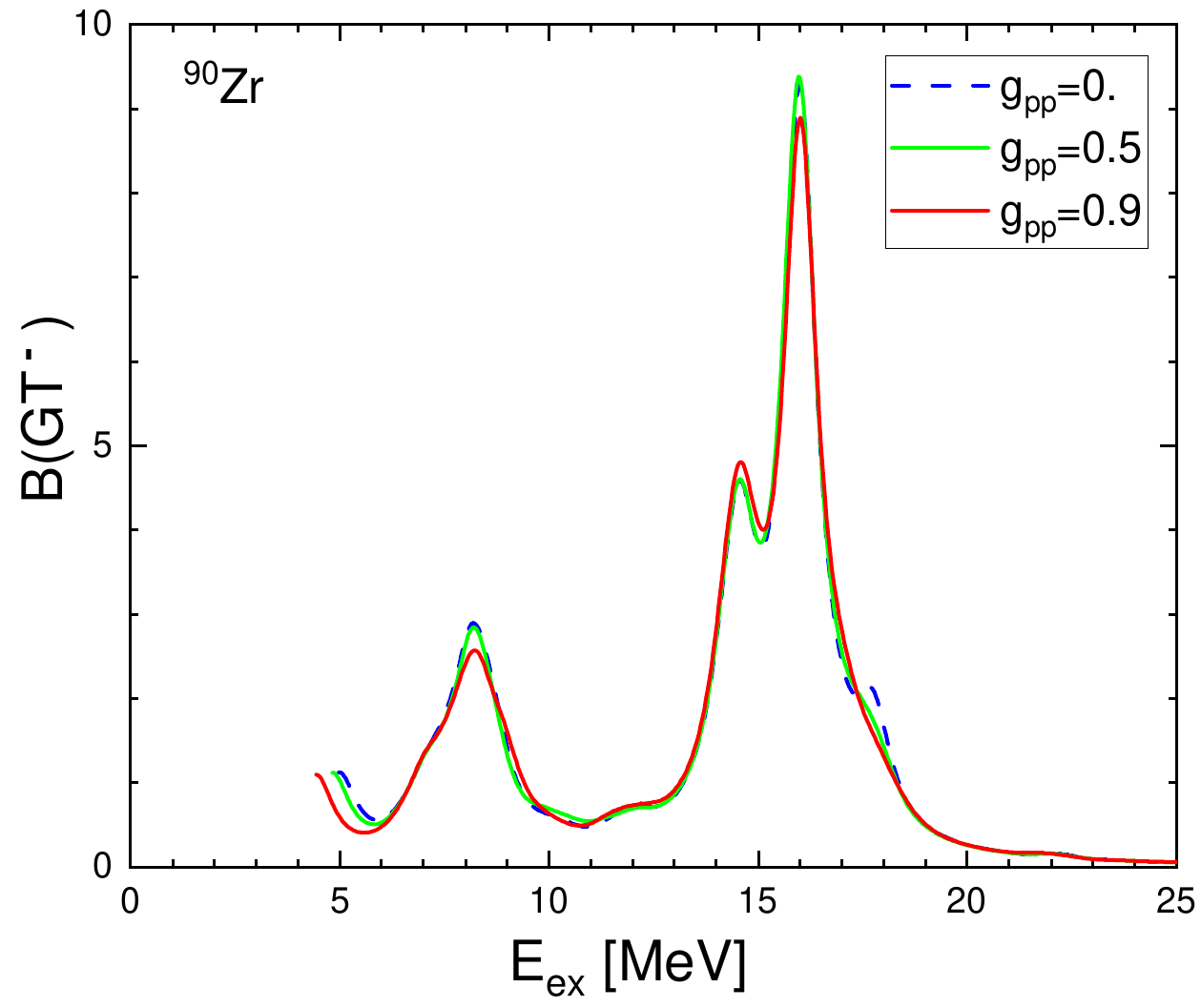}
\caption{(Color online) Effect of the $p-p$ interaction on the GT(-) transition strength distribution B(GT(-)) of $^{90}$Zr. Calculated results are obtained by changing the factor $g_{pp}$ as 0., 0.5, and 0.9 with a fixed normaization factor $g_{ph}$= 1.0.} \label{fig5}
\end{figure}

\begin{figure}
\centering
\includegraphics[width=0.49\linewidth]{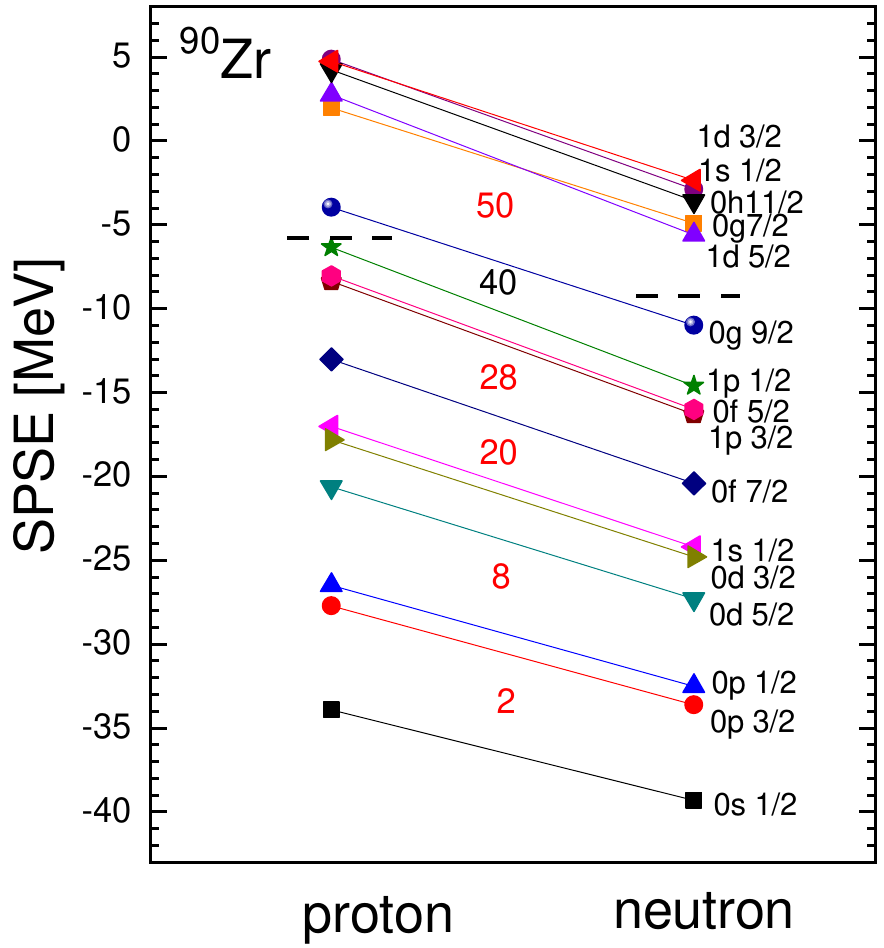}
\includegraphics[width=0.49\linewidth]{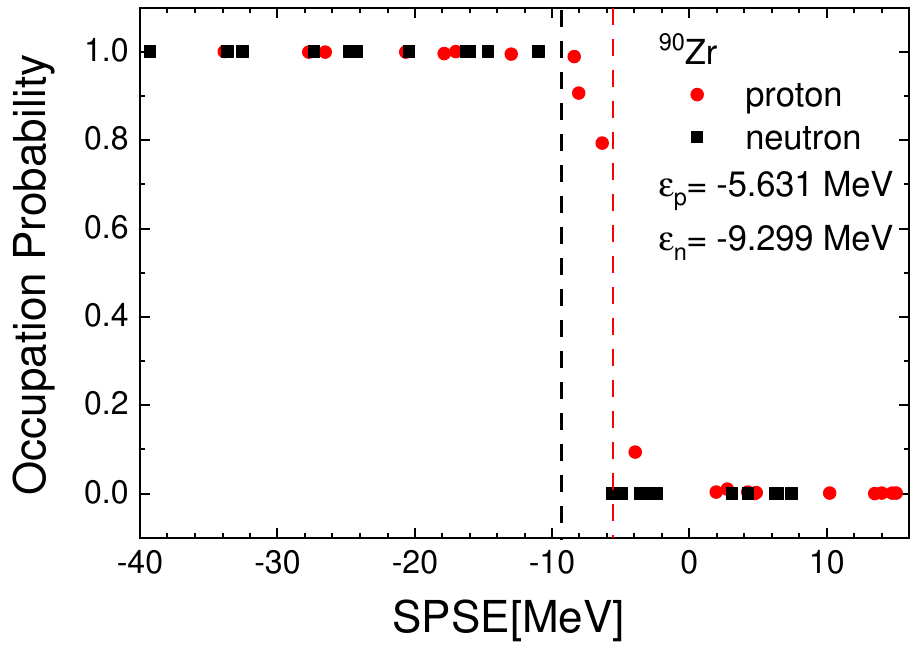}
\caption{(Color online) (left) SPS of $^{90}$Zr and (right) their occupation probabilities with their Fermi energies denoted as dashed lines.} \label{fig5-1}
\end{figure}

Since the $p-p$ interaction is almost negligible in the double magic nuclei as shown in Fig.\ref{fig3}, we investigated the property in $^{90}$Zr in Fig.\ref{fig5}, because the proton number $Z=$ 40 is not well-developed magic number.   
As displayed clearly in the right panel of Fig.\ref{fig5-1}, one can see the smearing of proton SPS in the vicinity of proton Fermi energy given in the left panel of Fig.\ref{fig5-1}.
Because of the melting of the Fermi surface, a slight difference is noticed in the B(GT$^-$) distribution by the change of $g_{pp}$ in Fig.\ref{fig5}, contrary to the negligible difference in the case of $^{48}$Ca in Fig.\ref{fig3}.

\begin{figure}
\centering
\includegraphics[width=0.65\linewidth]{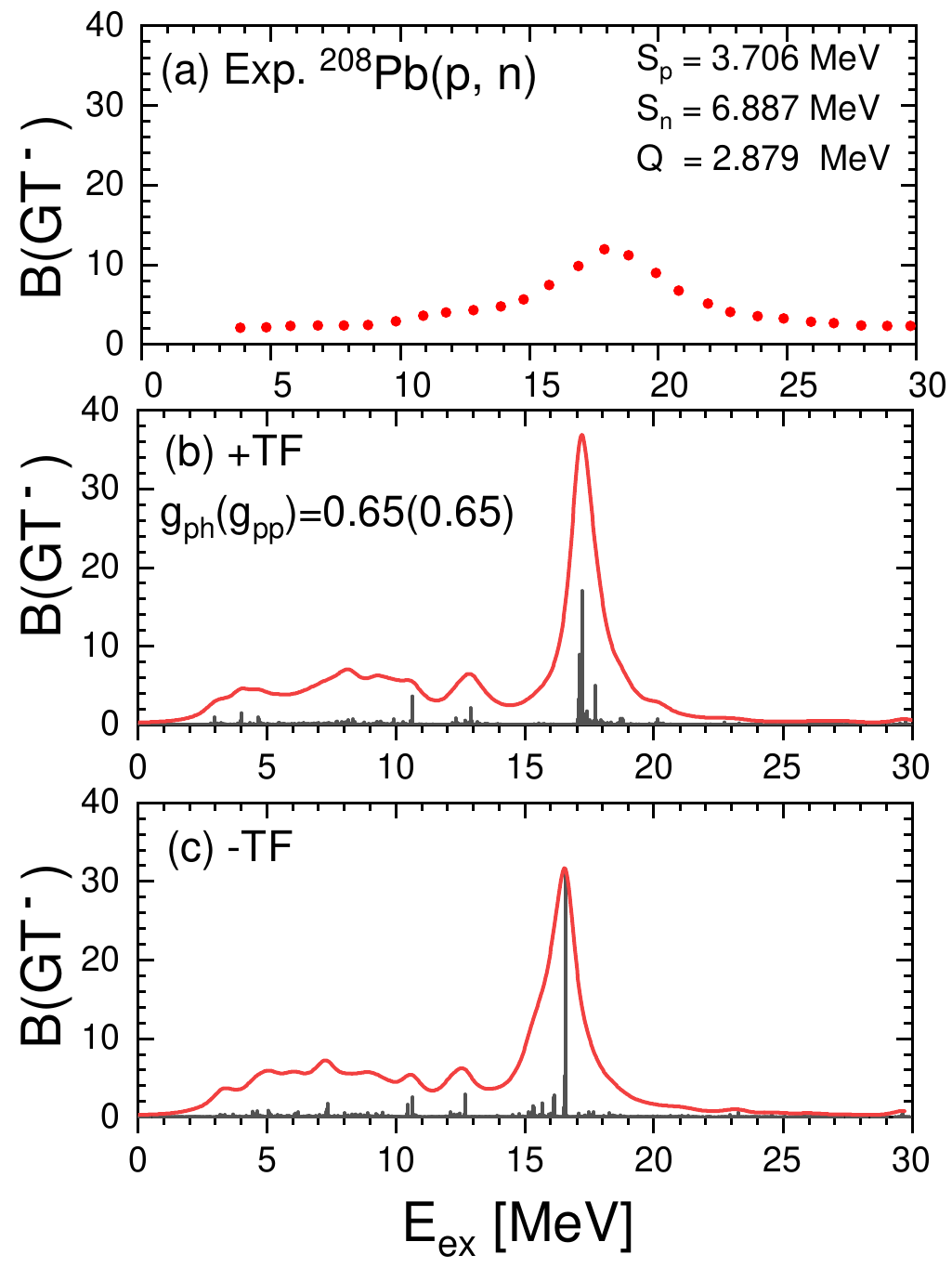}
\caption{(Color online) The same as Fig.\ref{fig4}, but for  $^{208}$Pb. The normalization factors are fixed to be $g_{ph}=$ 0.65 and  $g_{pp}=$ 0.65.
Experimental data by (p,n) reaction in panel (a) are from Ref. \cite{Wakasa12}.}
\label{fig6}
\end{figure}

In Fig.\ref{fig6}, we examine the TF effects on  a heavier double magic nucleus $^{208}$Pb, where one could not find the shoulder feature below the main GT peak in the experimental data
 as shown in Fig.\ref{fig6}(a),  in contrast to the other magic nuclei.    In the calculated results in Fig.\ref{fig6} (b) and (c), %the shoulder appears slightly, but
 the shoulder is smoothed out  below the main GR state, which has much larger width than the other nuclei.  This is due to the strong SO interactions for high $j$ orbits in the $p-h$ configurations and also the higher energy level density with the increase of the neutron number.  We notice that the main GT peak is shifted higher in energy by about 500 keV by the TF, while the entire spectrum is not much affected by the TF in $^{208}$Pb.

The effect of TFs on GT as well as spin-dipole (SD) excitations was discussed in a RPA model with Skyrme-type effective EDFs in Refs. \cite{Bai2011,Bai2009,Bai2009-2}.  In this RPA model, the triplet-even and triplet-odd TFs are
 introduced in both HF and RPA calculations.  The strengths of TFs are examined against the experimental data of GT and spin-dipole state in the closed shell nuclei.
 In GT case \cite{Bai2009}, the effect of triplet-even TF competes with that of triplet-odd one and shifts down the energy of GT states  contrary to the present result because of the different mean fields exploited.
 The best results of Skyrme-TFs are  obtained with an attractive triplet-even and a modest repulsive triplet-odd tensor forces which cancel each other largely and give a good agreement with the experimental excitation energies of GT states in $^{90}$Zr and $^{208}$Pb.

The GT strengths of closed shell nuclei have also been discussed with beyond mean field models; the particle-vibration coupling (PVC)+RPA model \cite{Niu2014,Niu2015}, the relativistic PVC model
 \cite{Robin2018}, and also subtracted second RPA (SSRPA) models \cite{Gambacurta2020,Yang2022}.  These beyond mean field models took into account more than $1p-1h$  model space and intended to describes not only the low-lying GT states, but also the main GT peaks.  Especially,  the calculated width of main GT peak \cite{Niu2014} is increased and becomes close to observed one by the inclusion of more than $1p-1h$ model space.   The calculated half-lives of double magic nuclei also show a very good agreement with experimental data \cite{Niu2015,Yang2023}.  In the SSRPA, the coupling to $2p-2h$
states is explicitly included and induces a large quenching of GT strength below E$_{ex} <$20 MeV which is consistent with the experimental observations.   In Refs. \cite{Gambacurta2020,Yang2022}, the Skyrme-type tensor interactions were introduced in  the SSRPA model.  Qualitatively, both the beyond mean field effect and the TF induced by Bonn potential improve the description of low-energy GT strength.    We found however that the precise description of GT energy and strength depends on the model adopted, for example,
the low-lying GT state in the $^{48}$Ca is well described in the present calculation, but the energy depends on the adopted TF in SSRPA \cite{Yang2022}.
 It is a future theoretical challenge to clarify  quantitative differences of the TF effect on the GT and SD states between the realistic TF derived from $G$-matrix  and the phenomenological ones.

Here we note that we did not use the Landau-Migdal approach, but we adopt the $G$-matrix from Br\"uckner HF
model. In principle, the effect of realistic $g_0’$ parameter of the spin-isospin channel is included in the BHF approach. For example, the
Landau-Migdal parameter was studied by using BHF model in Ref. \cite{Gamb2011} and shown to be consistent with phenomenological values adopted in experimental GT states analysis \cite{Suzu1999}.

\section{Summary and conclusions}

In summary,  we study the GT strength distribution of the magic nuclei $^{48}$Ca, $^{90}$Zr, $^{132}$Sn and $^{208}$Pb by the QRPA model with the realistic TF.
% the low energy GT peaks shift to high-lying GT states with the increase of the nuclear mass.
For $^{48}$Ca,
the low-lying GT strength appears clearly as a single peak both in the experiment and the calculated results. The attractive nature of tensor correlation for the  $p-h$ configuration with the same proton and neutron $j$ orbits shifts the peak position lower and gives a good agreement with the experimental data. For $^{90}$Zr,  %and $^{132}$Sn
 a shoulder appears below the main GT peak in the strength distribution,  and a plateau shows up  below the GT peak in $^{132}$Sn.  These GT strengths are slightly broader by the $p-h $ TF.
 On the other hand, the repulsive $p-h$ TF pushes the high-lying main GT peaks  up in energy  and makes  better agreements with  the experimental main GT peak positions.
This phenomenon is in contrast to the GT shift for $N=Z+2$ nuclei, where the attractive $p-p$ TF force shifts largely from the high-lying GT strength to the low-lying GT state, so called Low energy Super GT (LeSGT) state.
This difference  can be understood by the Pandya transformation, in  which the attractive $p-p$ tensor interaction in the $N=Z+2$ nuclei is changed to the repulsive $p-h$ interaction in the magic nuclei.

%In conclusion,
Thus, the TF property in the GT strength distribution of the magic nuclei shows up differently to  $N=Z+2$ nuclei, {\it i.e.}, it depends on whether the $p-p$ interaction is dominant or the $p-h$ interaction dominates the GT excited states. For the magic nuclei considered in this work, we showed that the attractive $p-h$ TF contributes to the evolution of the low-lying GT transition, while the repulsive $p-h$ TF contributes to  the high-lying GT excitation. By the competence of the TF to the spin-orbit force, the low-lying peak appears as a single peak for $^{48}$Ca, a left shoulder of the main GT peak for $^{90}$Zr and a plateau for $^{132}$Sn, and hindered in the background in a heavy nucleus $^{208}$Pb. The systematic study of the TF might be interesting including open-shell and deformed  nuclei. This study is planned for  a future project.

\section{Acknowledgement}
We appreciate APCTP for its hospitality during completion of this work. This work was supported by the National Research Foundation of Korea (Grant Nos. NRF-2018R1D1A1B05048026).  The work of MKC is supported by the National Research Foundation of Korea (Grant
Nos. NRF-2021R1A6A1A03043957 and NRF-2020R1A2C3006177).

\end{document}